\documentclass{article}
\usepackage[utf8]{inputenc}
\usepackage{authblk}
\usepackage{setspace}
\usepackage[margin=1.25in]{geometry}
\usepackage{graphicx}
\usepackage{subcaption}
\usepackage{amsmath}
\usepackage{lineno}
\usepackage{mathrsfs}
\usepackage{xcolor}
\usepackage{siunitx}

\usepackage{cleveref}

\crefformat{figure}{#1}

\DeclareUnicodeCharacter{0301}{\'{e}}

\usepackage[style=nejm, 
citestyle=numeric-comp,
sorting=none]{biblatex}
\title{Subcycle phase matching effects in short attosecond pulse trains}

\author[1*]{N Ouahioune}
\author[2,3]{R Martín-Hernández}
\author[1]{D Hoff}
\author[1]{PK Maroju}
\author[1]{C Guo}
\author[1]{R Weissenbilder}
\author[1]{S Mikaelsson}
\author[1]{A L'Huillier}
\author[4,5]{M Lucchini}
\author[1]{CL Arnold}
\author[1]{M Gisselbrecht}

\affil[1]{Department of Physics, Lund University, Lund, Sweden.}
\affil[2]{Grupo de Investigación en Aplicaciones del Láser y Fotónica, Departamento de Física Aplicada, Universidad de Salamanca, Salamanca, Spain.}
\affil[3]{Unidad de Excelencia en Luz y Materia Estructuradas (LUMES), Universidad de Salamanca, Salamanca, Spain.}
\affil[4]{Department of Physics, Politecnico di Milano, 20133 Milano, Italy.}
\affil[5]{Institute for Photonics and Nanotechnologies, IFN-CNR, 20133 Milano, Italy.}
\affil[*]{Address correspondence to: nedjma.ouahioune@fysik.lu.se}

\date{}

\begin{document}

\maketitle

\begin{abstract}
Attosecond pulses produced by High-order Harmonic Generation (HHG) in gases driven by intense laser fields have become a cornerstone technique for probing ultrafast electronic motion in matter. These applications require a good knowledge of the temporal and spectral properties of the emitted radiation. In this work, we generate a train of two to three attosecond pulses that we characterize using two-color laser-assisted photoionization. \textcolor{black}{An unexpected spectral behavior, with more pulses at high energies than at low energies, is observed when the carrier-to-envelope phase of the laser field is changed by 90$^\circ$.} HHG simulations indicate that the time-dependent phase matching of the harmonics 
contributes in a non-trivial way to the structure of the pulse train. Two-color laser-assisted photoionization enables us to unravel the dynamical influence of subcycle phase matching on the spectral properties of the attosecond pulse train, going beyond the predictions of the response of a single atom to a strong laser field.
\end{abstract}

\section{Introduction}
The generation of high-order harmonics of an intense laser in gas targets \cite{FerrayJPhysB_1988,McPhersonJOSAB_1987} has enabled the production of attosecond pulses of light in the extreme-ultraviolet (XUV) range \cite{PaulScience_2001,HentschelNature_2001}. The process depends on the highly non-linear interaction of the intense laser with an isolated atom (or molecule) - a microscopic effect known as the single-atom response \cite{KulanderProceedings_1993,LewensteinPRA_1994,CorkumPRL_1993}, and the in-phase propagation of the generated XUV radiation and the driving laser field - a macroscopic effect called phase matching \cite{LHuillierPRA_1992,BalcouPRA_1997,WeissenbilderNatRevPhys_2022}. In general, the properties of the attosecond pulses depend on the interplay between the single-atom response and propagation effects, which influences the efficiency, coherence, and spatiotemporal structure of the XUV emission \cite{KazamiasPRL_2003, FerrariNatPhot_2010, SansoneScience_2006, WikmarkPNAS_2019}. The impact of reshaping of the fundamental field on HHG has also been pointed out for high laser intensities \cite{GaardeOL_2006,MajorJPhysPhotonics_2020,MajorFiOLS_2024,KretschmarSciAdv_2024,VismarraLSA_2024}.

\begin{figure}[ht]
    \centering
    \includegraphics[scale=0.32]{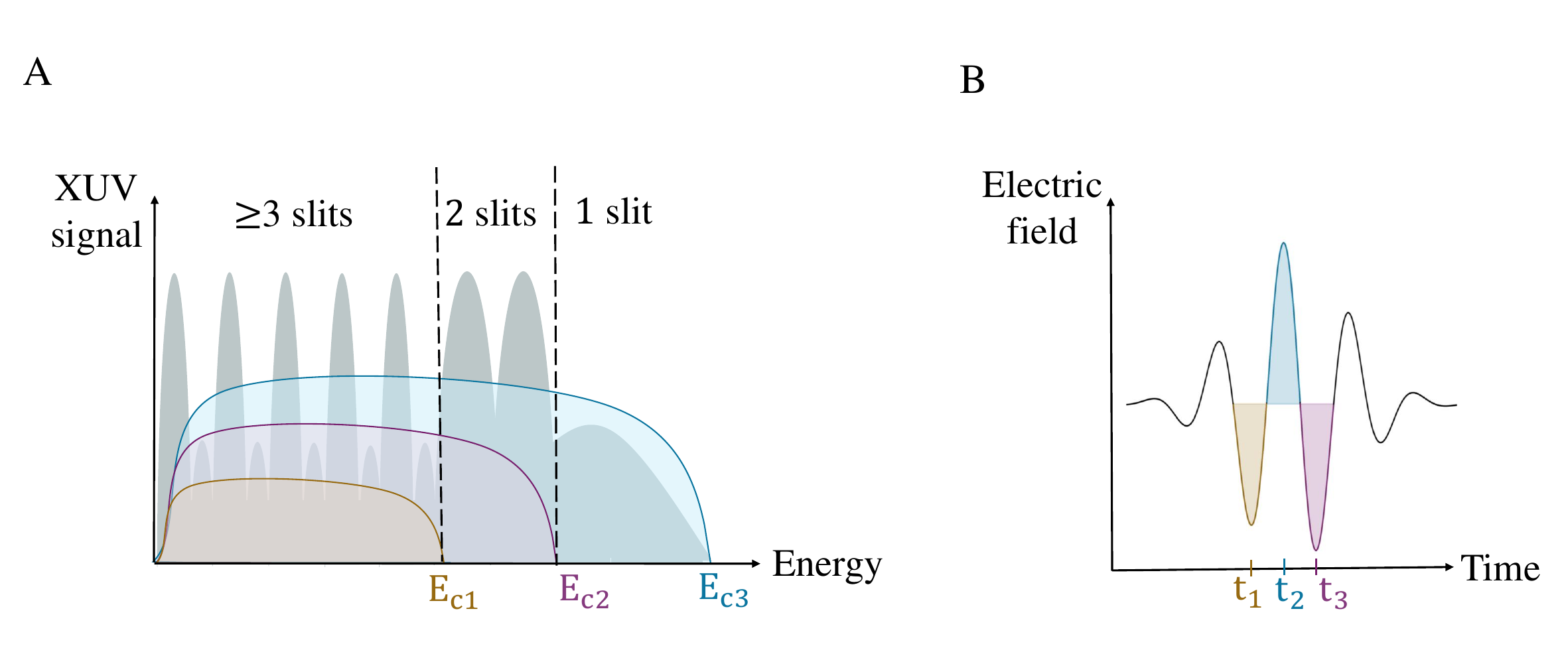}
    \caption{(A) Illustration of the spectral content of attosecond pulses (temporal slits) generated in each half-cycle of a short laser pulse, shown in (B). Each color (purple, blue, and orange) refers to a specific pulse, with a unique cut-off energy $E_c$. The pattern resulting from the interference is shown in gray.}
    \label{fig:schematic_intro}
\end{figure}

The generation of high-order harmonics is due to the interference of attosecond pulses (temporal slits) created in the subcycle light-matter interaction \cite{ChengPNAS_2020}, similar to the emission of electrons in above-threshold ionization (ATI), due to the interference of attosecond electron wavepackets \cite{LindnerPRL_2005}. In HHG, the harmonics are separated by twice the laser photon energy while in ATI, the detection breaks the inversion symmetry, and electron energy peaks are separated by the laser photon energy. When the laser pulse is short ($\leq$ $10$ fs), the number of slits is small and the spectral range of the attosecond pulses differ from one half-cycle to the next. In the single-atom response, this spectral range can be obtained simply from the cut-off law \cite{CorkumPRL_1993,KrausePRL_1992}, as shown schematically in Fig.~\ref{fig:schematic_intro}.

Attosecond pulses can be temporally characterized using laser-assisted photoionization schemes such as streaking \cite{SansoneScience_2006,HentschelNature_2001,GoulielmakisScience_2008} and RABBIT (Reconstruction of Attosecond Beating by Interference of Two-photon transitions) \cite{PaulScience_2001,MairesseScience_2003}. In these schemes, electron spectra are measured as a function of the delay between the XUV radiation and a phase-locked infrared (IR) field. In streaking, the electron energy oscillates and the temporal properties can be obtained from the amplitude and contrast of these oscillations \cite{MairesssePRA_2005}. In RABBIT, the absorption/emission of an additional IR photon leads to additional peaks, called sidebands, and the temporal properties of an average attosecond pulse in a train can be obtained from the phase of the sideband oscillations \cite{PaulScience_2001, LopezMartensPRL_2005}. In both cases, retrieval methods taking advantage of the entire spectrum as a function of delay have been successfully developed \cite{MairesssePRA_2005,LucchiniOE_2015,LucchiniAppSci_2018, DolsoAPLPhotonics_2023, KeathleyNewJPhys_2016,OrfanosAPLPhotonics_2019}.

In this work, we create attosecond pulse trains (APT) using few-cycle laser pulses, with a stable carrier-to-envelope phase (CEP), focused in a gas of argon atoms. Due to the short laser duration ($\leq$6 fs), the APTs contain a few attosecond pulses \cite{ChengPNAS_2020} and their number depends on the CEP of the driving laser. We characterize these APTs by laser-assisted photoionization of helium using a three-dimensional momentum spectrometer \cite{MikaelssonJNanophotonics_2021}. For certain CEP values, the spectrograms, which are the electron energy spectra as a function of delay, exhibit features that cannot be explained by the structure of the APTs predicted by the single-atom response alone. This is confirmed by a time-frequency analysis of the APTs retrieved from the experimental data \cite{LucchiniOE_2015, KeathleyNewJPhys_2016}. Supported by 3D simulations and a 1D-model, we show that the temporal properties of the APTs are also influenced by time-dependent phase matching in the nonlinear medium, with, in some cases, a non-trivial number of pulses as a function of the XUV energy. This experimental and theoretical study emphasizes the importance of subcycle phase matching for the accurate prediction and manipulation of the properties of attosecond light pulses.

\section{Methods}

\begin{figure}[ht]
    \centering
    \includegraphics[scale=0.45]{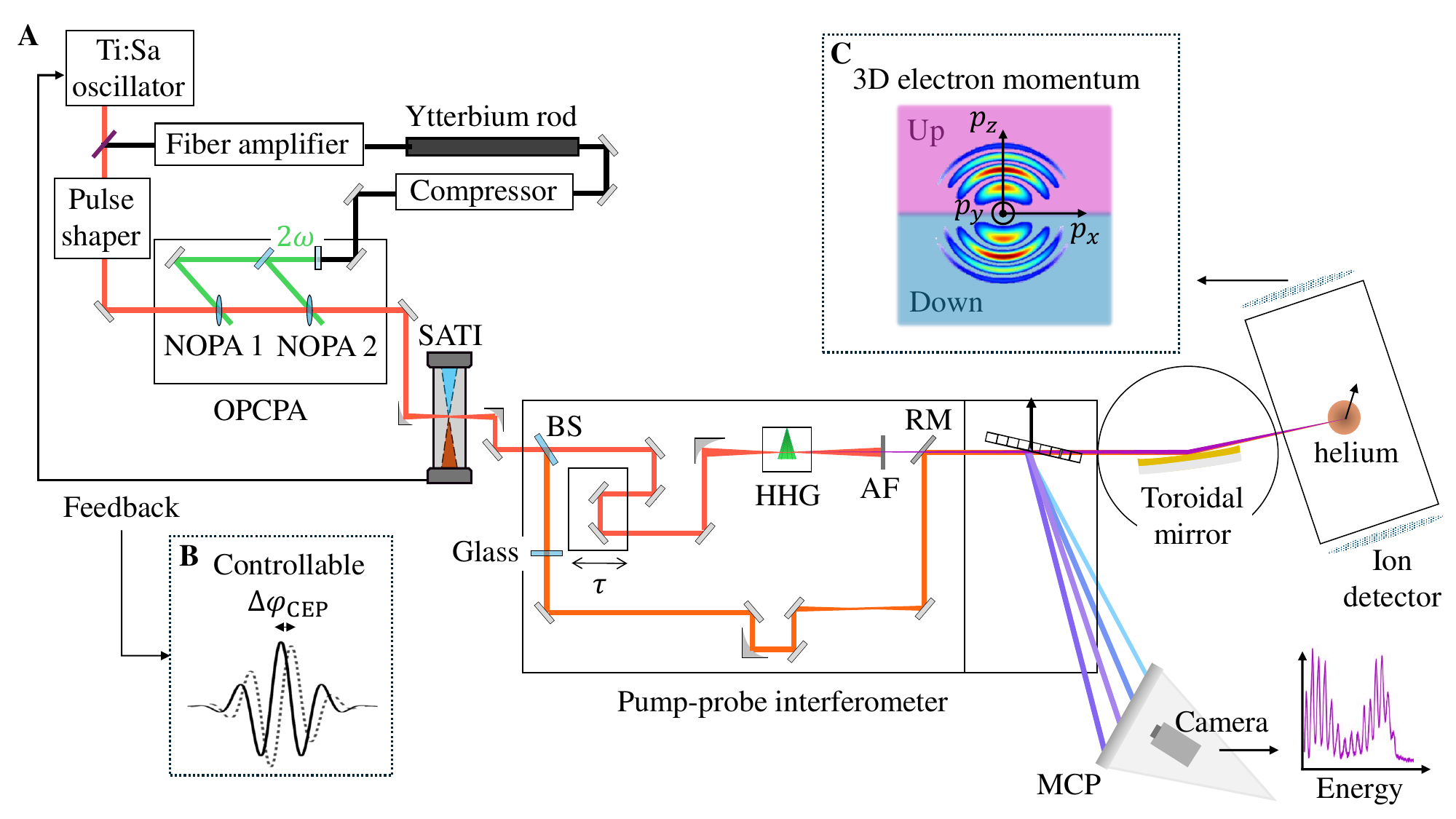}
    \caption{(A) Schematic of the experimental setup. Laser pulses from a Ti:Sapphire oscillator are amplified using two nonlinear optical parametric amplification (NOPA) stages. The relative CEP of the output pulses is controlled shot-to-shot through the feedback of a stereo Above-Threshold-Ionization (SATI) device to a wedge pair in the oscillator \cite{WittmannNatPhys_2009}. The CEP-controlled pulses are separated into a probe arm and a variably delayed pump arm by a beam splitter (BS). Short attosecond pulse trains are produced by the IR driving field using HHG and recombined with the probe before being focused into the sensitive region of a 3D electron spectrometer.
    (B) The controllable relative CEP $\Delta \varphi_{\mathrm{CEP}}$ of the $\leq$ 6 fs laser pulses. (C) Momentum projected along $p_x$ and $p_z$ of the photoemitted electrons. The electrons can be distinguished between those emitted with $p_z>0$ (``Up") and those emitted with $p_z<0$ (``Down"), where $z$ is the axis along the detector and of the polarization of the light. BS: Beam splitter; SATI: Stereo-ATI; AF: Aluminium filter; RM: Recombination mirror; MCP: Micro-channel plate.}
    \label{fig:exp_schematic}
\end{figure}

The experiment, shown in Fig.~\ref{fig:exp_schematic}, was performed using an optical parametric chirped pulse amplification (OPCPA) system, providing sub-6-fs long pulses at a central wavelength of 850 nm with up to \SI{10}{-\micro\joule} pulse energy at a repetition rate of 200 kHz \cite{MikaelssonJNanophotonics_2021}. A stereo-ATI (SATI) setup installed in the beam path stabilized the CEP on a shot-to-shot basis, providing a stability of around 160 mrad (corresponding to a precision of $10^\circ$) \cite{WittmannNatPhys_2009}. Furthermore, the CEP was controlled by a wedge pair in the oscillator (not shown in the schematic). The laser beam was then split into a pump (XUV) and a probe (IR) arm by a beamsplitter (BS). In the pump arm, the beam was tightly focused (beam waist $w_0 =$ \SI{5}{\micro\meter}\textcolor{black}{, Rayleigh length $z_R\sim$ \SI{100}{\micro\meter})} into a thin argon gas jet ($\sim$ \SI{36}{\micro\meter} length) at a pressure of $4$ bar, reaching an intensity of $\sim 1.5\times10^{14}$ W/cm$^2$ to generate a comb of phase-locked odd-order harmonics. The low-order harmonics, below $\sim$15 eV, and the generating field were filtered out with an aluminum foil (AF) of 200-nm. The XUV pump was recombined with the probe using a holey mirror (RM). The delay $\tau$ between the XUV and the IR in the probe arm can be controlled using a delay stage in the pump arm. Once recombined, both fields were focused by a toroidal mirror into a helium gas target; the probe intensity was estimated to be $\sim 10^{11}$ W/cm$^2$. A reaction microscope was used to measure the three-dimensional momentum of the photoelectrons created by the absorption of the XUV-only or the XUV+IR radiation. In our experiments, the delay between the XUV light and the IR probe was scanned for different CEP settings. Changing the CEP allowed us to vary the number of pulses in the train, typically between two and four \cite{ChengPNAS_2020}.

\section{Results and Discussion}
\begin{figure}[ht]
    \centering
    \includegraphics[scale=0.4]{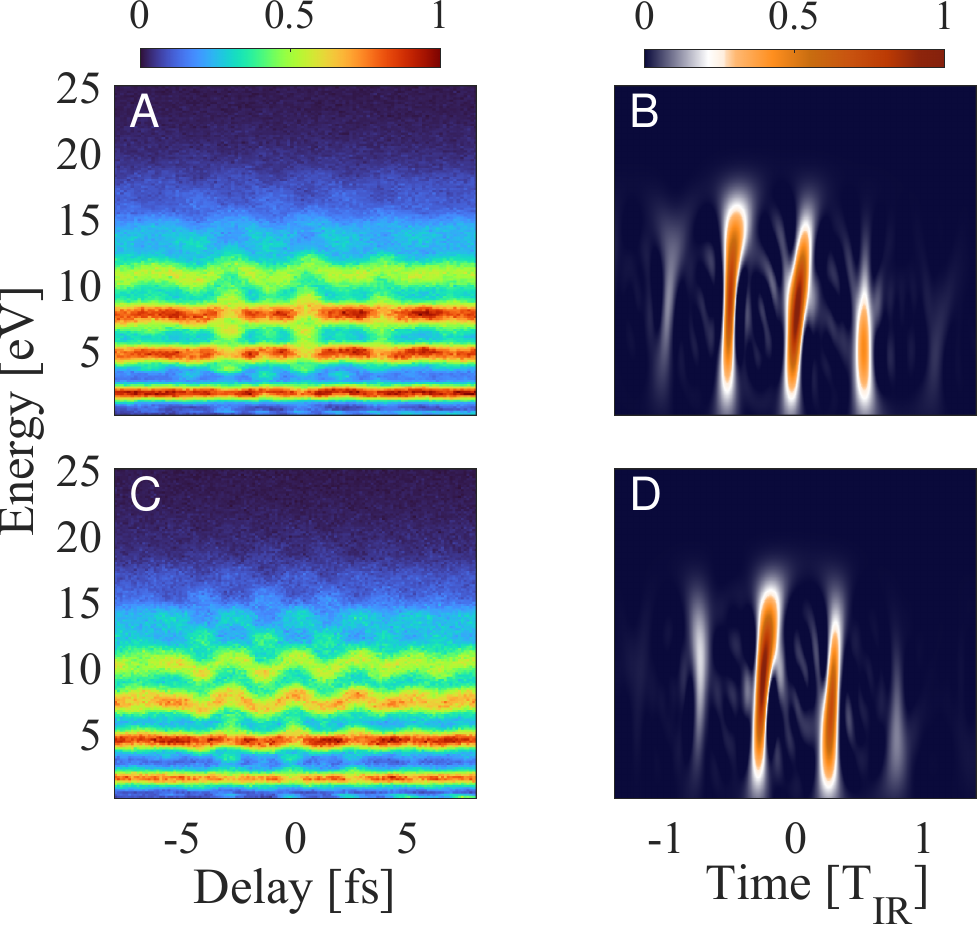}
    \caption{(A and C) Experimental photoelectron spectra as a function of the XUV-IR delay for two different CEP values separated by 90$^\circ$. Only electrons photoemitted in the upper hemisphere around the polarization axis of the light fields are shown. (B and D) Wigner distributions of the attosecond pulses retrieved from the experimental spectrograms through the refined extended ptychographic iterative engine (rePIE) \cite{LucchiniAppSci_2018}.}
    \label{fig:fig_3}
\end{figure}

Figures \ref{fig:fig_3}A and C present photoelectron spectra as a function of the delay for two different CEP values separated by 90$^{\circ}$. In the figure, only electrons emitted in a solid angle of $+ 2\pi$ sr (``up'') along the common polarization axis of XUV and IR fields are shown (see Fig.~\ref{fig:exp_schematic})~\cite{ChengPNAS_2020}. At large delays, when the fields do not overlap, the photoelectron peaks are separated by a kinetic energy of $2\hbar\omega$ ($\hbar$ is the reduced Planck constant and $\omega$ is the laser frequency) due to single-photon ionization by the comb of harmonics. At delays where the XUV and the IR overlap, additional features form, which depend on the kinetic energy and the CEP.  Above 10 eV in Fig.~\ref{fig:fig_3}A, and between 7 and 10 eV in Fig.~\ref{fig:fig_3}C, the kinetic energy of the photoelectrons oscillates as a function of the delay with a periodicity equal to the laser period $\sim$2.6 fs. In other spectral regions, additional electron peaks are observed at kinetic energies between those due to the absorption of consecutive harmonics. In particular, in Fig.~\ref{fig:fig_3}C, the pattern looks like a ``chessboard'' \textcolor{black}{over a large energy range from 12 eV to the highest observed energy}. This remarkable energy dependence comes from the small number of attosecond pulses in the train. 
Oscillations arise from the interference between two electron wave packets, similar to a double-slit experiment \cite{ChengPNAS_2020}. The additional electron peaks are due to interference of more than two electron wave packets, leading to sidebands, similar to RABBIT measurements performed with multiple-cycle pulses~\cite{PaulScience_2001, MairesseScience_2003}. 

\textcolor{black}{In the supplementary material, we also show the spectra obtained by collecting electrons emitted in the opposite direction (``down'') in Fig.~\cref{fig:Figure_S1}. The results are very similar to Fig.~\ref{fig:fig_3}A,C, except that the positions of the maxima and minima of the oscillations are shifted by half a laser period \cite{ChengPNAS_2020}. Consequently, collecting electrons over the full solid angle removes the CEP-dependent features. We note that the behavior of the photoelectron spectra do not vary much with angle in each hemisphere.}

In the single atom response, illustrated in Fig.~\ref{fig:schematic_intro}, the minimum driving field intensity $I_{\mathrm{min}}(q)$ required for a harmonic order $q$ to reach the plateau region can be obtained from the cut-off law, $q\hbar\omega= I_p+ 3.17 U_p$, where $I_p$ is the ionization energy of argon and $U_p$ is the ponderomotive energy equal to $\alpha h I/(m_e\omega^2)$. In this expression, $\alpha$ is the fine structure constant, $I$ the laser intensity and $m_e$ the electron mass. This intensity, expressed as 
  \begin{equation}
   I_{\mathrm{min}}(q)=\frac{m_e \omega^2}{3.17 \alpha h}(q\hbar\omega-I_p),
  \end{equation}
increases linearly as a function of the harmonic order $q$. For our few-cycle pulses ($\leq$ 6 fs), this implies that the highest-order harmonics are generated only during a few half-cycles of the electric field at the center of the pulse. Consequently, the number of attosecond pulses in the train depends on a chosen spectral range. It decreases as the central photon energy increases. With our ultrashort pulses, only two half-cycles (or less) reach the intensity for the generation of the highest orders. On the other hand, lower-order harmonics are generated over many half-cycles. The above argument would always result in RABBIT-like features at low photoelectron kinetic energies and an oscillatory pattern (streaking-like pattern) at higher photoelectron energies, for any CEP. From these arguments
, the experimental observation in Fig.~\ref{fig:fig_3}C, cannot be explained.

To understand the temporal structure of the attosecond pulse trains, we apply the refined extended pthycographic iterative engine (rePIE) \cite{LucchiniOE_2015,LucchiniAppSci_2018,KeathleyNewJPhys_2016}, described in the supplemental material (see supplementary material). The retrieved spectrograms are in good agreement with the experimental ones, as shown in Fig.~\cref{fig:Figure_S2} of the supplementary material. We describe the properties of the APTs using the Wigner-Distribution (WD), defined as :
\begin{equation}
    W(t,\Omega) =\! \int_{-\infty}^{+\infty}\!\! \! \mathscr{E}_{\mathrm{APT}}\left(t+\tfrac{\tau}{2}\right)  \mathscr{E}^*_{\mathrm{APT}}\left(t-\tfrac{\tau}{2}\right) e^{-i \Omega \tau} \,d\tau,
    \label{eq:WD}
\end{equation}
where $\Omega$ is the XUV frequency and $\mathscr{E}_{\mathrm{APT}}$ is the XUV field. The advantage of using the WD compared to other time-frequency representations is that the spectral and temporal resolution is only limited by the uncertainty principle.
Figures \ref{fig:fig_3}B,D shows the Wigner representations for the APTs retrieved from the experimental results in Fig.\ref{fig:fig_3}A,C. Figure \ref{fig:fig_3}B shows essentially three pulses at low energies and two pulses above 12 eV. In comparison, Fig.~\ref{fig:fig_3}D shows two dominating pulses up to 10 eV and three pulses at higher energies. The observed spectral variation of the structure of the APT cannot be explained by only considering the single-atom response.

\begin{figure}[t]
    \centering
    \includegraphics[scale=0.4]{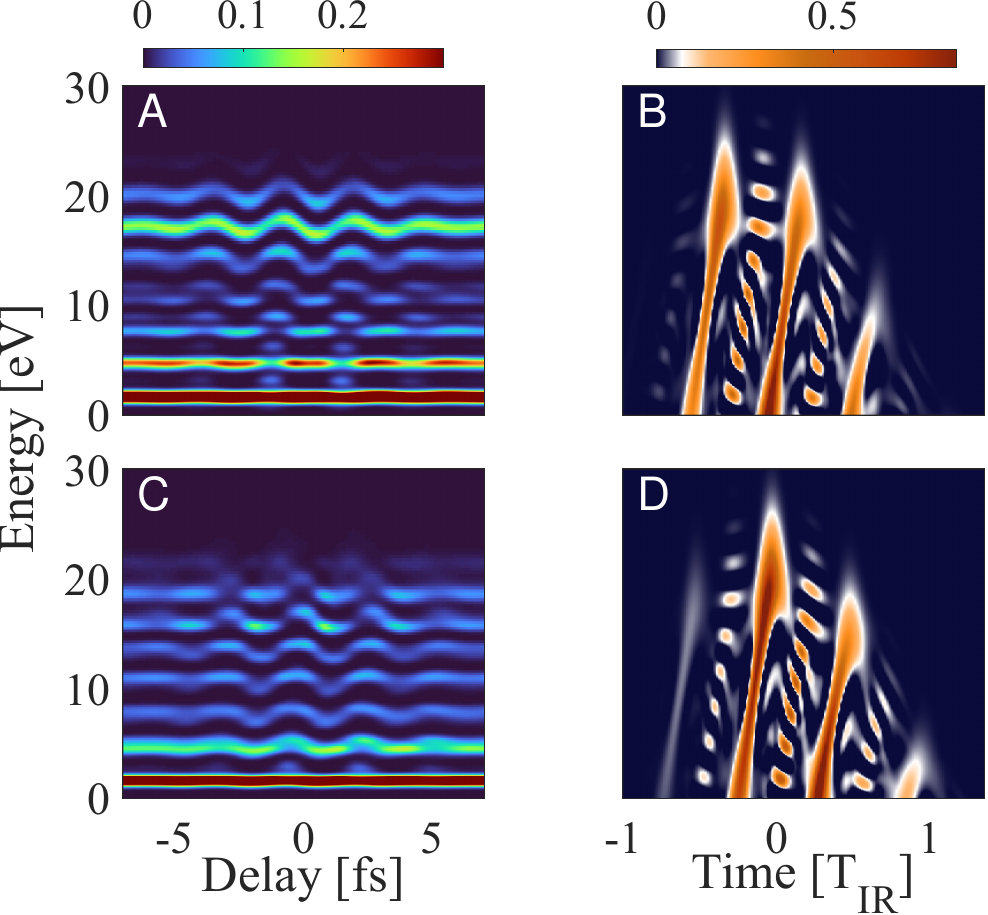}
    \caption{(A and C) Photoionization spectrograms as a function of the kinetic energy of the electrons and of the XUV-IR delay simulated using the 3D model. (B and D) Wigner representation of the simulated attosecond pulse trains. \textcolor{black}{The CEPs are (A and B) $70^\circ$ and (C and D) $160^\circ$.}}
    \label{fig:fig_4}
\end{figure}

We performed advanced HHG simulations that account for both microscopic and macroscopic effects. Specifically, we compute the far-field harmonic emission by solving the Schr\"odinger equation within the SFA \cite{LewensteinPRA_1994} and by including the propagation and absorption of the fundamental and harmonic fields in a thin nonlinear medium while accounting for the longitudinal phase matching of the XUV radiation \cite{Hernandez-GarciaPRA_2010}. \textcolor{black}{This calculation goes beyond the slowly-varying envelope approximation, including in particular subcycle ionization dynamics with the Yudin-Ivanov CEP-dependent model described in \cite{Yudin_Ivanov_etafe}.} The driving laser field is defined as $E(t) =E_0 f(t) \exp[ i ( \omega t + \varphi_{\mathrm{CEP}} ) ] $, where $E_0$ is the amplitude of the electric field, $f(t)=\sin^2{( \pi t/t_{\mathrm{p}} )}$ (for $0 \leq t \leq t_{\mathrm{p}}$) and $\varphi_{\mathrm{CEP}}$ is the CEP. We perform a systematic study of HHG as a function of pressure, intensity, pulse duration and focusing conditions. We find that our observations are well reproduced for the parameters corresponding to the experimental ones (see experimental section). An exception is the pulse duration which is chosen to be 4.85 fs, slightly less than the experimental estimation ($\sim 6$ fs). \textcolor{black}{In these conditions, the ionization degree is relatively small, typically a few percent. Ground-state depletion is negligible, and the driving laser is not significantly reshaped during propagation. The single-atom response can be considered to be uniform throughout propagation in the gas, see Section 3 of the supplementary material}. The calculated APTs are then used to simulate laser-assisted photoionization spectra using SFA. The results shown in Fig.~\ref{fig:fig_4}A,C, \textcolor{black}{using CEPs of $70^\circ$ and $160^\circ$}, reproduce the experimental trends. The Wigner representations in Fig.~\ref{fig:fig_4}B,D are similar to those obtained from the retrieved APTs (see Fig.~\ref{fig:fig_3}), showing  the same number of pulses across the different energy regions and an overall good agreement with the experiment. A noticeable difference is the tilt observed in the simulation due to the attosecond chirp \cite{MairesseScience_2003}. The absence of tilt in the experiment likely originates from the aluminum filter, which compensates for the attosecond chirp \cite{LopezMartensPRL_2005}. \textcolor{black}{The effect of the spectral phase variation due to photoionization is here negligible.} These theoretical results confirm that CEP-dependent macroscopic effects influence in a non-trivial way the spectral content of the individual attosecond pulses.

To gain insight into the influence of macroscopic effects on the pulse train, we developed a one-dimensional model of time-dependent harmonic phase matching, including subcycle variations. The total phase mismatch $\Delta k(q,t)$ depends on the medium properties and on the laser parameters and can be written as \cite{WeissenbilderNatRevPhys_2022}
\begin{equation}
    \Delta k(q,t) \! =\! \Delta k_{\mathrm{foc}}(q) \! + \! \Delta k_{\mathrm{at}}(q) \! + \! \Delta k_{\mathrm{dip}}(q,t) \! + \! \Delta k_{\mathrm{fe}}(q,t) 
\end{equation}
where $\Delta k_{\mathrm{foc}}(q) = - q / z_\mathrm{r}$ comes from the laser focusing (through the variation of the Gouy phase over the Rayleigh length $z_\mathrm{r}$), $\Delta k_{\mathrm{at}}(q)$ is due to the dispersion of the neutral medium, $\Delta k_{\mathrm{dip}}(q,t)$ comes from the dipole phase with the time-dependent intensity taking into account only the short trajectory contribution, $\Delta k_{\mathrm{fe}}(q,t)\propto - q \rho\eta_{\mathrm{fe}}(t)$ is due to the free electrons and depends on the time-dependent ionization degree $\eta_{\mathrm{fe}}(t)$. The ionization degree is often estimated using a cycle-averaged approximation \cite{KeldyshSovPhysJETP_1965,PerelomovSovPhysJETP_1966,WeissenbilderNatRevPhys_2022}. Here, we account for the subcycle variation of the ionization degree, allowing us to include CEP effects. In addition, we neglect the CEP-slip across the focus as the medium length is small compared to the Rayleigh length.  

\begin{figure}[ht]
    \centering
   \includegraphics[scale=0.3]{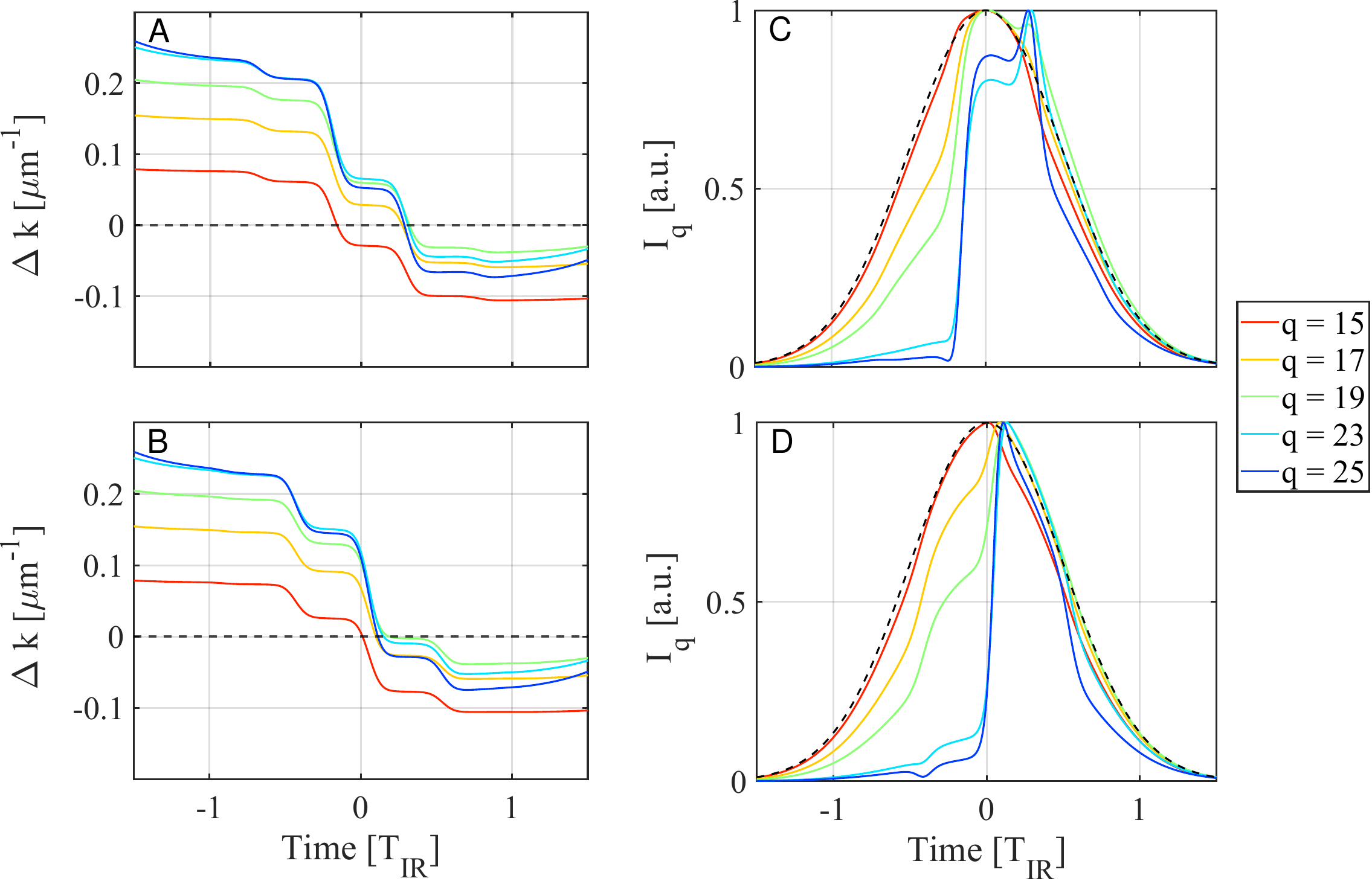}
    \caption{\textcolor{black}{(A and B) Total phase mismatch as a function of time in units of the laser period ($T_\mathrm{IR}$) for four harmonic orders $q$ for the CEPs corresponding to Fig.~\ref{fig:fig_3}B. The horizontal black dashed line corresponds to perfect phase matching. (C and D) Corresponding yield for each harmonic order $q$. The CEPs are 70$^\circ$ (top) and 160$^\circ$ (bottom). Dashed lines indicate the temporal profile of the single-atom response.}}
    \label{fig:fig_5}
\end{figure}

\textcolor{black}{Figures \ref{fig:fig_5}A and C show the temporal variation of $\Delta k(q,t)$ for five harmonic orders and two CEPs equal to 70$^\circ$ and 160$^\circ$}. $\Delta k(q,t)$ strongly varies with time and for the different orders. The steps reflect the variation of $\Delta k_{\mathrm{fe}}(q,t)$ through the ionization rate. The difference in heights between the curves comes from the $q$-dependence of the atomic dispersion $\Delta k_{\mathrm{at}}(q)$ and the dipole phase $\Delta k_{\mathrm{dip}}(q,t)$. Perfect phase matching ($\Delta k =0$) is in general achieved only during a short time, less than one laser period, leading to temporal confinement of the harmonic emission. This time interval, in the conditions examined in the present work, moves to a later time as the harmonic order increases. \textcolor{black}{At a CEP of 70$^\circ$, phase matching, corresponding to $|\Delta k(q,t)|\leq$ \SI{0.03}{\micro\meter}$^{-1}$, is achieved simultaneously for all harmonics. At a CEP of 160$^\circ$, it is achieved at a later time and differently for harmonics 17-21 and harmonics 23-25.}

For on-axis propagation in a uniform gas density, the intensity of the $q$th harmonic field at the exit of the medium can be expressed as \cite{ConstantPRL_1999,RuchonNewJPhys_2008}
\begin{equation}
    I(q,t)\! \propto \! [\rho d(q,t)]^2  \frac{\cosh{\left[ \kappa_q L \right]} - \cos{\! \left[ \Delta k(q,t)L \right ] }  }{\kappa_q^2 \!+\! \Delta k^2(q,t)} e^{-\kappa_qL},\!
    \label{eq:Iq}
\end{equation}
where $d(q,t)$ is the dipole moment obtained by using an effective power law dependence ($|d|^2\propto I^{2.6}$) \cite{WeissenbilderNatRevPhys_2022}, $\rho \, = \, 8.8$ kg/m$^3$ is the experimental gas density, $L=$ \SI{36}{\micro\meter} is the medium length and $\kappa_q$ is the absorption coefficient \cite{SamsonJESRP_2002}. \textcolor{black}{The laser spatio-temporal properties are assumed constant throughout the interaction length since reshaping effects were found negligible (see supplementary material). The spectrograms simulated using this model, shown in Fig.~\cref{fig:Figure_S4} of the supplementary material, reproduce the experimental observations, in contrast to those obtained with the single-atom response (see Fig.~\cref{fig:Figure_S5}).} 

Figures \ref{fig:fig_5}\textcolor{black}{B and D} present the harmonic intensity as a function of time for different harmonic orders and \textcolor{black}{CEPs equal to 70$^\circ$ and 160$^\circ$}. In this model, the single-atom response leads to a temporal dependence $\propto I^{2.6}(t)$ for all harmonics, shown in dashed lines, while the inclusion of phase matching adds a subcycle $q$-dependent modulation of the temporal profile following the variation of $\Delta k(q,t)$ (see Fig.~\ref{fig:fig_5}A \textcolor{black}{and C}). \textcolor{black}{We observe that phase matching leads to a temporal confinement of the high-order harmonics. At the CEP of 160$^\circ$, harmonics 23 and 25 are generated after the maximum intensity of the driving pulse and temporally confined to less than a half-cycle. In contrast, at 70$^\circ$, all the harmonics appear before the maximum intensity, and are generated over a longer time. CEP-dependent phase matching} changes the relative strength and duration of the attosecond pulses, depending on the spectral region, as observed in our experimental and theoretical results (see Fig.~\ref{fig:fig_3}-\ref{fig:fig_4}). \textcolor{black}{In XUV far field spectra, variations in the number of slits can be observed with the pressure, as shown in Fig.~\ref{fig:Figure_S6} of the supplementary material.} While this model provides a valuable insight into the influence of phase matching on the spectro-temporal structure of the attosecond pulses, it cannot describe the details of the complex three-dimensional spatio-temporal dynamics of the generation process.

\section{Conclusion}
\textcolor{black}{In conclusion, we show that photoionization by short attosecond pulse trains in the presence of a dressing field leads to CEP-dependent photoelectron spectra. Our study emphasizes the sensitivity of laser-assisted photoionization for retrieving the intricate time-frequency dependence of short attosecond pulse trains. The behavior of photoelectron spectra can be explained by subcycle phase matching, which leads to temporal confinement of some harmonics, thereby acting as a passive pulse shaper. The spectral amplitude of each attosecond pulse in the train, therefore, varies in a non-trivial way, depending on the CEP. This work provides a comprehensive experimental and theoretical insight into subcycle phase-matching dynamics for short attosecond pulse trains.}

\section*{Acknowledgments}
We thank E. Constant for fruitful discussions.

\subsection*{Author Contributions} 
 N.~O. analyzed the data and conceived the 1D model. R.~M.~H. performed the 3D simulation. D.~H. and  S.~M. performed the experiments. P.~K.~M. performed complementary analysis. C.~G, C.~L.~A., and S.~M. designed and constructed the light sources. M.~L. performed the retrieval of the experimental data. P.~K.~M., R.~W., C.~L.~A., C.~G., A.~L. and M.~G. contributed to the scientific discussion.  M.~G. supervised the work. N.~O. wrote the manuscript with inputs from A.~L. and M.~G. All authors provided feedback to the manuscript.

\subsection*{Funding}
M.~G., A.~L. and C.L.~A acknowledge support from the Swedish Research Council (Grant Nos. 2020-05200, 2023-04603, 2021-04691) and Alice Wallenberg Foundation. A.~L. acknowledges support from the European Research Council (Grants No. 884900, No 851201) and the Knut and Alice Wallenberg Foundation through the Wallenberg Centre for Quantum Technology (WACQT). This project has also received funding from the Ministerio de Ciencia e Innovación (grant PID2022-142340NB-I00) and from the Department of Education of the Junta de Castilla y León and FEDER Funds (Escalera de Excelencia CLU-2023-1-02 and grant No. SA108P24).

\subsection*{Competing interests}
The authors declare that there is no conflict of interest regarding the publication of this article.

\subsection*{Data Availability}
Data underlying the results presented in this paper are not publicly available at this time but may be obtained from the authors upon reasonable request.


\section{Supplementary material}

\setcounter{figure}{0}
\renewcommand{\thefigure}{S\arabic{figure}}

\textcolor{black}{\subsection{Angular-dependence of the photoelectron spectra}}

\begin{figure}[ht]
    \centering
    \includegraphics[scale=0.25]{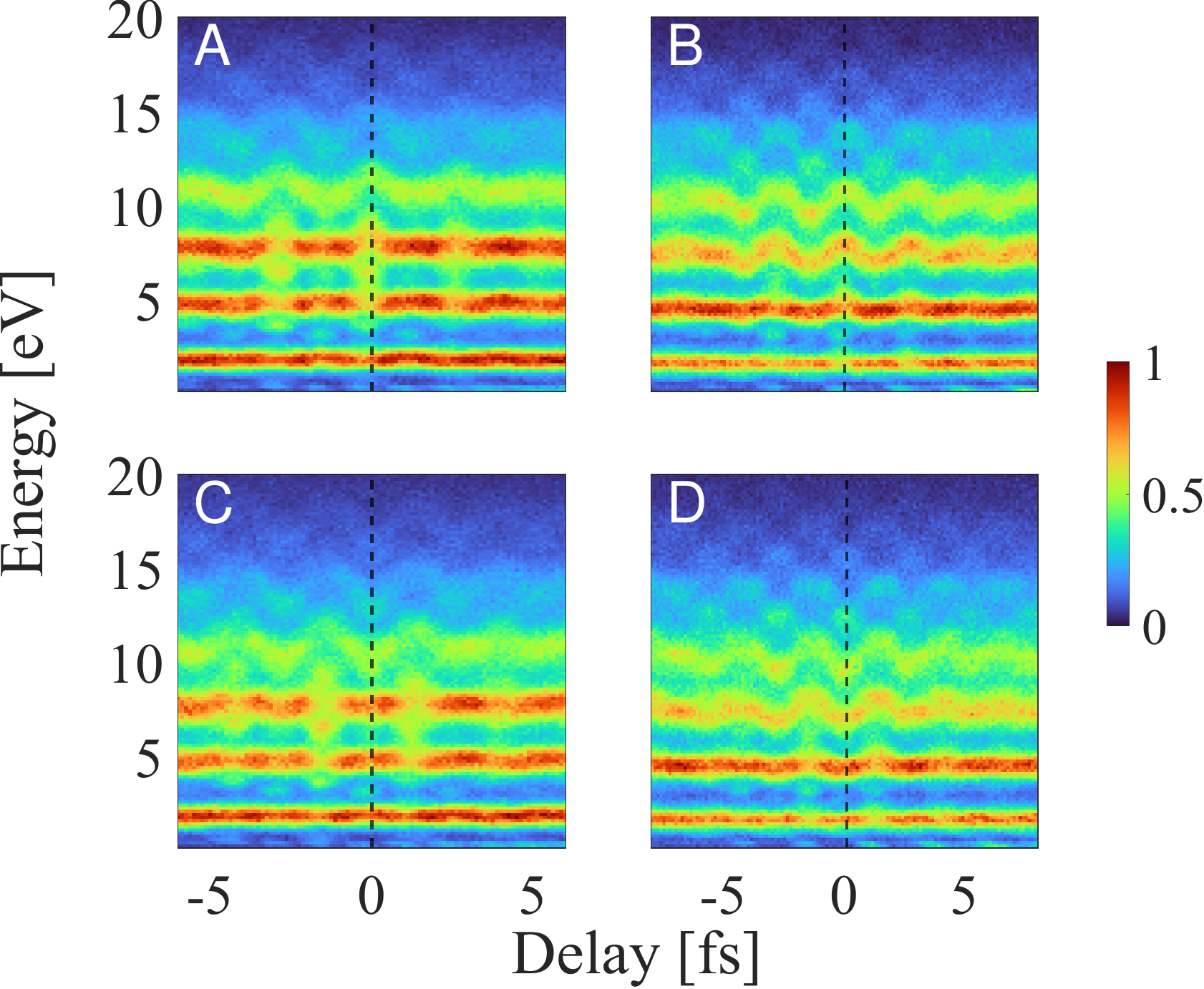}
    \caption{\textcolor{black}{Experimental photoelectron spectra for electrons emitted in the (A and B) upper ($p_z>0$) and (C and D) lower ($p_z<0$) hemisphere of the detector. } }
    \label{fig:Figure_S1}
\end{figure}

\textcolor{black}{The presented results include only electrons emitted in the upper hemisphere $p_z>0$  (toward the detector) shown in Fig.~\cref{fig:Figure_S1}A,B. In Fig.~\cref{fig:Figure_S1}C,D, the electrons emitted for $p_z<0$ (in the opposite direction) are shown. The results are very similar, except that the positions of the maxima and minima of the oscillations are shifted by half a laser period, see vertical dashed lines. In the results presented here, obtained in helium, the electron distribution does not vary significantly within each half hemisphere.}

\subsection{Pulse reconstruction from the experimental data}
The experimental spectrograms are reconstructed using the refined extended ptychographic iterative engine (ePIE) \cite{LucchiniAppSci_2018}. In this approach, exploiting the central-momentum approximation (CMA), a first quick reconstruction is obtained with the extended ptychographic iterative engine (ePIE) \cite{LucchiniOE_2015}. The ePIE output becomes then the starting guess for a second step based on the “Volkov transform” generalized projections algorithm (VTGPA) \cite{KeathleyNewJPhys_2016}. This later does not contain the CMA, but it is typically more expensive in terms of computational cost and time. The initial ePIE step, though, assures that good convergence is reached after few iterations (50-100), thus removing the CMA from the output while reducing overall the reconstruction time. It is important to stress that in our case, since the photoelectron spectra extends towards zero eV, rePIE assures that the low energies are correctly reconstructed. Most of the more common algorithms, based on CMA would instead fail.\\

\begin{figure}[ht]
    \centering
    \includegraphics[scale=0.35]{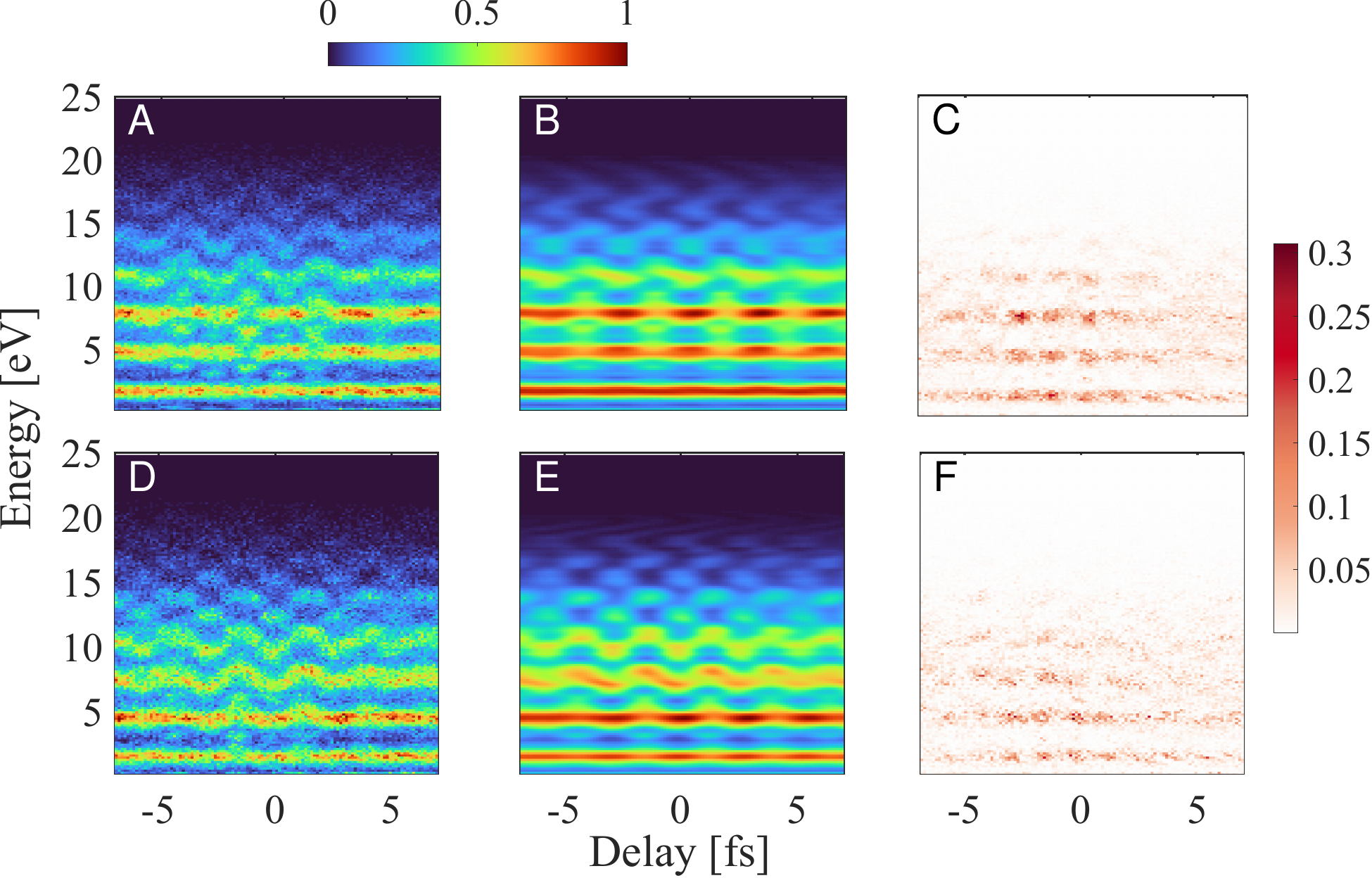}
    \caption{(A and D) Experimental and (B and E) retrieved spectrograms. (C and F) Root mean square difference between the retrieved and the input spectrograms.}
    \label{fig:Figure_S2}
\end{figure}

The output of the rePIE reconstructions are shown in Fig.~\cref{fig:Figure_S2}. To minimize the effect of the angular averaging done by the detector, the retrieval is performed on photoelectrons emitted in a small solid angle (total aperture of $+$20$^\circ$ around the polarization axis of the light fields). For the initial ePIE step the experimental traces have been resampled so that the delay and energy axes have a spacing of 0.3 fs and 12 meV. The resulting time resolution of the reconstructed pulses is 83 as. The code starts from a random guess and reaches good convergence after 2500 iterations, yielding a FROG error (normalized r.m.s. distance between the experimental and reconstructed traces \cite{MurariOE_2020}) of $4.9 \times 10^{-2}$ and $4.5 \times 10^{-2}$ for CEP1 and CEP2, respectively. The second VTGPA step runs over the native experimental spectrogram (delay step of 120 as and energy step of 167 eV). The IR envelope is described with 7 anchors while its phase is a polynomial of the 5th order. The resulting time resolution for the reconstructed pulses is 30 as. Good convergence is reached after 50 iterations leading to a FROG error of $4.4 \times 10^{-3}$ for CEP1 and $3.2 \times 10^{-3}$ for CEP2.\\

We note that to prove the applicability of rePIE under the experimental conditions used in this work, we tested the algorithm against the simulated spectrogram. The spectro/temporal properties of the XUV are fully reconstructed for both CEP values, proving the applicability of our approach.

\begin{figure}[ht]
    \centering
    \includegraphics[scale=0.3]{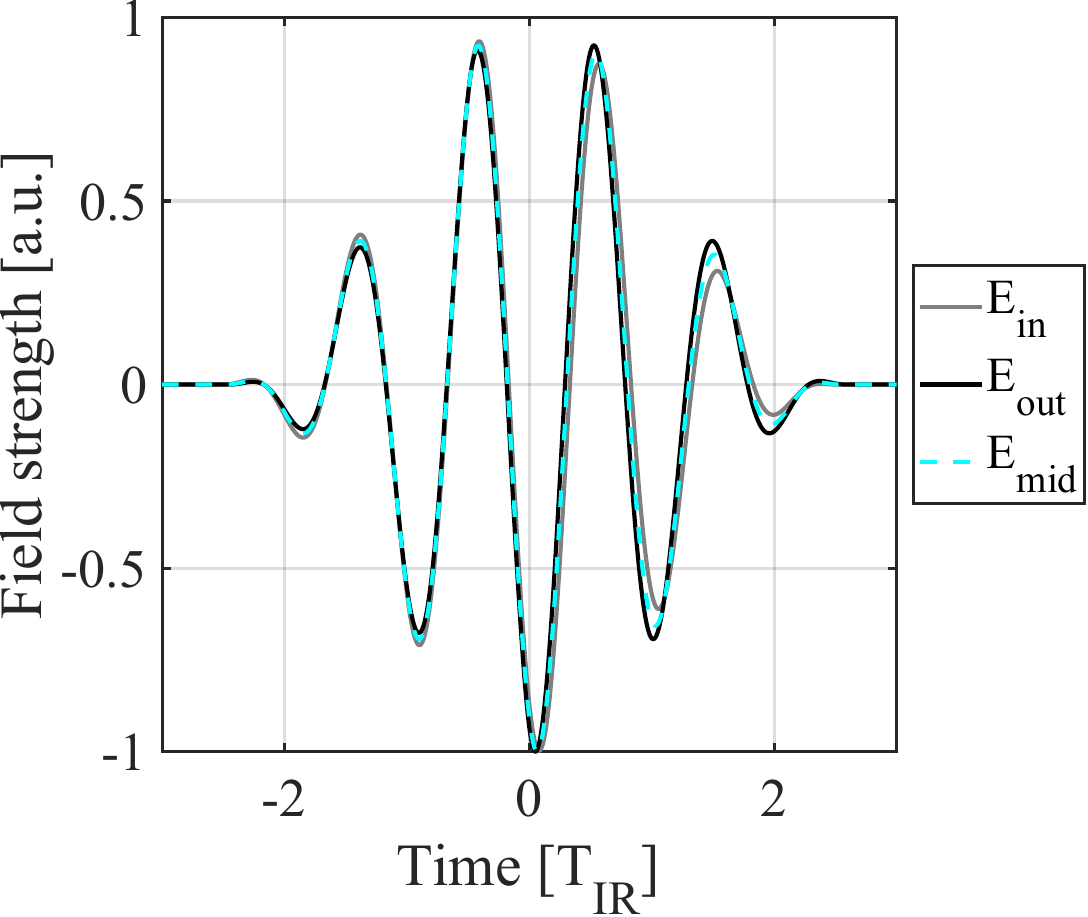}
    \caption{\textcolor{black}{Electric field at the entrance and exit of the HHG medium for a CEP of 160$^\circ$.}}
    \label{fig:Figure_S3}
\end{figure}

\textcolor{black}{\subsection{Reshaping of the driving field}
In the three-dimensional simulations, the reshaping of the electric field (plasma and gas dispersion) during propagation is included. At each position, the electric field is updated and the single-atom response computed using the strong-field approximation. In our experimental conditions, consisting of moderately small peak intensities and a thin medium, the electric field at the input $E_{\mathrm{in}}$, middle $E_{\mathrm{mid}}$ and output $E_{\mathrm{out}}$ of the gas target remains unchanged, as depicted for example in Fig.~\cref{fig:Figure_S3}. The change in phase due to focusing (Gouy phase) does not change the electric field throughout the medium. As a consequence, the single-atom response is uniform throughout the interaction length.}

\subsection{One-dimensional temporal phase matching model}
\begin{figure}[ht]
    \centering
    \includegraphics[scale=0.35]{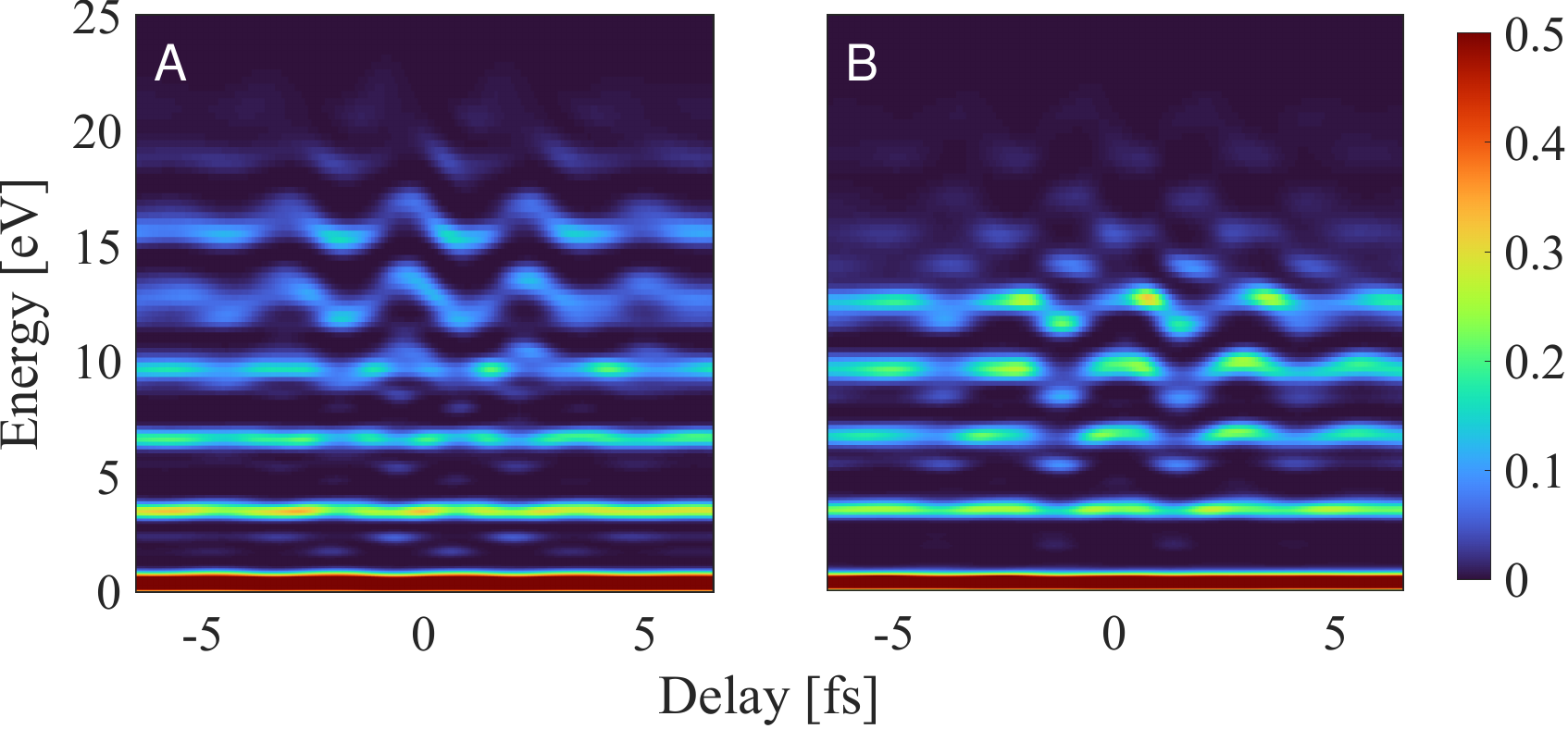}
    \caption{\textcolor{black}{Spectrograms obtained with SFA simulations using the APTs calculated with the 1D analytical model. The CEPs are (A) 70$^\circ$ and (B) 160$^\circ$.}}
    \label{fig:Figure_S4}
\end{figure}

Using the 1D analytical model presented in the main text, we show in Fig.~\cref{fig:Figure_S4} that the experimental spectrograms can be well-reproduced, in particular, the non-trivial variation of the number of slits with the energy.

\begin{figure}[ht]
    \centering
    \includegraphics[scale=0.35]{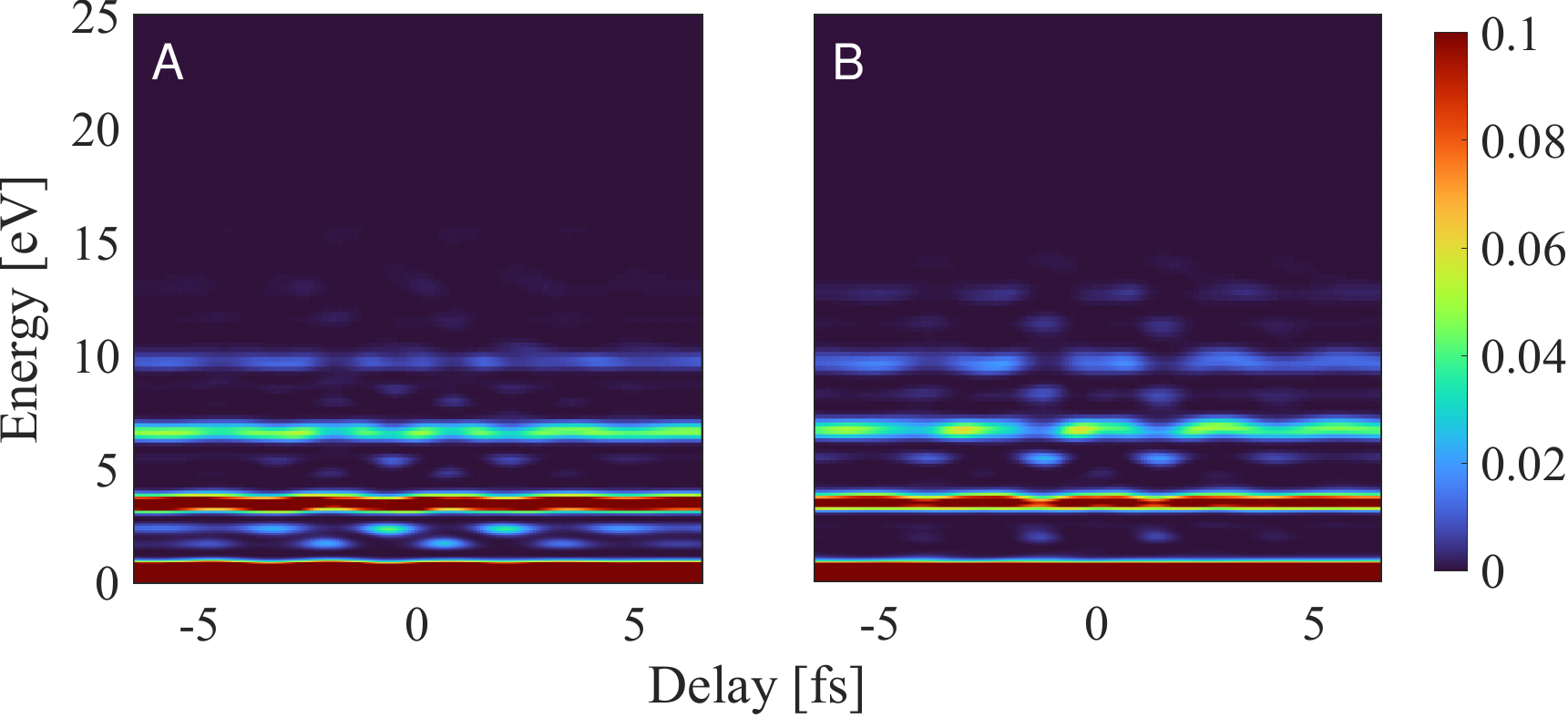}
    \caption{\textcolor{black}{Spectrograms obtained using the single-atom response calculated with the strong-field approximation. The CEPs are (A) 70$^\circ$ and (B) 160$^\circ$.}}
    \label{fig:Figure_S5}
\end{figure}

\textcolor{black}{The spectrograms simulated using the single-atom response for the same laser parameters are presented in Fig.~\cref{fig:Figure_S5}. A RABBIT-like pattern (3-4 slits) is observed at all energies, since attosecond pulses produced in the rising edge of the driving field have a significant contribution.}

\begin{figure}[ht]
    \centering
    \includegraphics[scale=0.45]{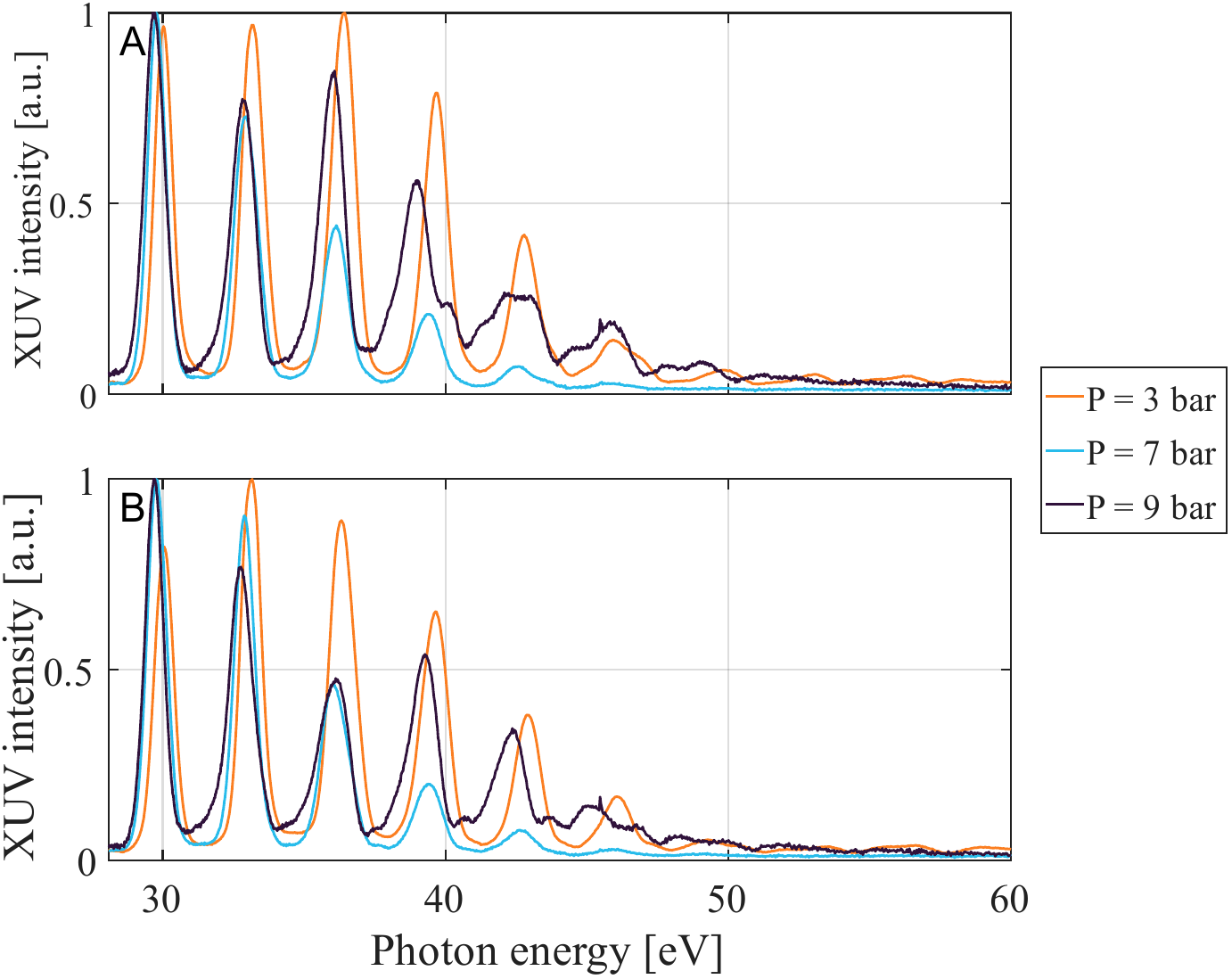}
    \caption{\textcolor{black}{Far field XUV signal integrated radially as a function of the energy for two CEPs separated by 90$^\circ$. The different colors indicate different backing pressures. The curves are normalized to their maximum.}}
    \label{fig:Figure_S6}
\end{figure}

\textcolor{black}{\subsection{Far field harmonic spectra}
Complementary experiments were carried out to study the effect of pressure on the harmonic spectra, for similar laser parameters and focusing geometry as the results presented in the article. In Fig.~\cref{fig:Figure_S6}, the far-field XUV spectra for two CEP values separated by 90$^\circ$ are shown at different backing pressures, which is about twice higher than the pressure in the jet \cite{MikaelssonJNanophotonics_2021}. The spectral width of harmonics generated at low pressures (3 and 7 bar) is independent of the harmonic order. When the pressure is increased to 9 bar, however, harmonics at high energies become broader, indicating that harmonics are generated during a shorter time corresponding to a smaller number of attosecond pulses. Figure ~\cref{fig:Figure_S6}A shows broadband peaks compatible with generation of two attosecond pulses while Fig.~\cref{fig:Figure_S6}B shows peaks in-between the odd orders, which suggests the presence of three attosecond pulses in the high-energy region. This confirms that macroscopic effects can change the number of attosecond pulses in selective spectral regions.}

\printbibliography

\end{document}